\documentclass[useAMS,usenatbib]{mn2e}
\usepackage{graphicx}

\usepackage{times}
\usepackage[dvips]{color}

\def\mmm{$(m-M)_0$}
\def\ebv{$E(B-V)$~}

\def\nodata{ }

\def\gsim{\;\lower.6ex\hbox{$\sim$}\kern-7.75pt\raise.65ex\hbox{$>$}\;}
\def\lsim{\;\lower.6ex\hbox{$\sim$}\kern-7.75pt\raise.65ex\hbox{$<$}\;}

\newcommand{\rr}{${\mathcal R}$}

\title[The old open cluster Berkeley 32]
{The old anticentre open cluster Berkeley 32: membership and fundamental
parameters\thanks{ Based on observations made with the Italian
Telescopio Nazionale Galileo (TNG) operated on the island of La Palma by
the Fundaci\'on Galileo Galilei of the INAF (Istituto Nazionale di
Astrofisica) at the Spanish Observatorio del Roque de los Muchachos of
the Instituto de Astrofisica de Canarias.}}
\author[D'Orazi et al.]
{V. D'Orazi$^{1}$\thanks{E-mail: valentina.dorazi@studio.unibo.it (VDO),
angela.bragaglia@bo.astro.it (AB), monica.tosi@bo.astro.it (MT), 
difabrizio@tng.iac.es (LDF), held@pd.astro.it (EVH)},
A. Bragaglia$^{2}$, 
M. Tosi$^{2}$, 
L. Di Fabrizio$^{3}$ and 
E. V. Held$^{4}$\\
\\
$^{1}$ Dipartimento di Astronomia, Universit\`a di Bologna, via Ranzani 1, 
40127 Bologna (Italy)\\
$^{2}$ INAF-Osservatorio Astronomico di Bologna, via Ranzani 1, 40127 Bologna
(Italy)\\
$^{3}$  Fundaci\'on Galileo Galilei - INAF
Calle Alvarez de Abreu 70, 38700 Santa Cruz de La Palma, TF (Spain)\\
$^{4}$ INAF-Osservatorio Astronomico di Padova, 
vicolo dell'Osservatorio 5, 35122 Padova (Italy)\\
}

\begin{document}

\date{ }

%\pagerange{\pageref{firstpage}--\pageref{lastpage}} \pubyear{2002}

\maketitle

\label{firstpage}

\begin{abstract}
We have obtained medium-low resolution spectroscopy and $BVI$ CCD imaging of
Berkeley~32,  an old open cluster  which lies in the anticentre direction. 
From the radial velocities of 48 stars in the cluster direction we found that
 31 of them, in crucial evolutionary phases, are probable  cluster members, with
an average radial velocity of +106.7 ($\sigma$ = 8.5) km s$^{-1}$.

From isochrone fitting  to the  colour magnitude diagrams of Berkeley~32 we
have obtained an age of 6.3 Gyr, \mmm = 12.48 and \ebv = 0.10. The best fit is
obtained with Z=0.008.
A consistent distance, \mmm~$\simeq 12.6 \pm 0.1$, has been derived from the 
mean magnitude of  red clump stars with confirmed membership; we
may assume \mmm~$\simeq 12.55 \pm 0.1$

The colour magnitude diagram of the nearby field 
observed to check for field stars contamination 
 looks intriguingly similar to that of the Canis Major overdensity.

\end{abstract}

\begin{keywords}
Open clusters and associations: general --
open clusters and associations: individual: Berkeley 32 --
Hertzsprung-Russell (HR) diagram

\end{keywords}

\section{Introduction}

Open clusters (OCs) are an important population of the disc of our Galaxy and
may be used to trace its properties (e.g., spiral structure, kinematics, age
and age distribution, metallicity, etc; see \citealt{friel95} for a review). In
particular, by studying the old component of the OC population we may be able
to trace the disc properties at all epochs since its formation, and to derive
its evolution. The number of known -- and well studied -- old OCs has  steadily
grown in recent years; what we still miss is a populous sample analyzed in a
complete and homogeneous way to derive the cluster properties like membership,
age, distance, reddening, and (detailed) chemical abundances using both
photometric and low- and high-resolution spectroscopic data.

We have therefore undertaken a project devoted to  precise and homogeneous
derivation of the fundamental properties of a large and well chosen sample of
old open clusters. The goal of our project is to get insight on the formation
and chemical enrichment mechanisms of the Galactic disc. A description of the
motivations of our  program, and a status report of its photometric part are
given in \cite{bt06}.

As part of this project, we present here a study of   Berkeley~32 (Be~32), an
old anticentre open cluster, with coordinates $\alpha_{2000}=06^h58^m07^s$,
$\delta_{2000}=+06^\circ 25'43''$, and $l=208^\circ$,  $b=+4.4^\circ$. We have
obtained medium resolution spectra of about 50 stars to derive the membership,
as we have recently done for  Berkeley 29 \citep*{bht05}, while $B, V, I$
CCD images are used to derive age, distance, reddening, and approximate
metallicity.

The cluster has already been studied by several authors. \cite{scott95} 
obtained low-resolution spectra and measured radial velocities of 14 stars,
finding that 10 of them were cluster members and obtaining an average velocity
of 101 km~s$^{-1}$, with an error of 10 km~s$^{-1}$. These  velocities were
adopted also by \cite{friel02}, who determined metallicities - with an
individual error of about 0.1 dex - and found an average  [Fe/H] = $-0.5$. 

\cite{km91} obtained  $U, B, V$ and Washington photometry of Be~32; combining
different methods, like the Morphological Age Ratio (see Sect. 3), the
comparison to well studied clusters like NGC~188 and M~67, and main sequence
fitting, they determined that Be~32 has an age of about 6 Gyr, a metallicity
[Fe/H] $= -0.37 \pm 0.05$, reddening \ebv = 0.16, and distance modulus $(m-M)_V
= 12.95 \pm 0.15$.  
\cite{rs01}, working with $V, I$ data, reached similar conclusions; they
employed isochrone fitting and red clump luminosity to derive an 
age of 6.3 Gyr, [Fe/H] $\sim -0.20$, \ebv = 0.08, and true
distance modulus $(m-M)_0 = 12.60 \pm 0.15$. 
Finally,  \cite{pasj04} observed Be~32 (that they indicate with the alternative
name of Biurakan 8) as part of a survey of 14  anticentre clusters; their $B,
V, I$ photometry is of somewhat lower quality, and results they derive from
isochrone fitting and morphological parameter indicators are not in perfect
agreement, but they confirm that Be~32 is among the oldest OCs.

The paper is organized as follows: observations and reductions are described in
Section 2, the CMDs are presented in  Section 3, together with results on
isochrone fitting;  the radial velocities and memberships are derived in
Section 4; a summary can be found in Section 5.

\begin{figure}
 \includegraphics[bb=65 150 550 635, clip, width=84mm]{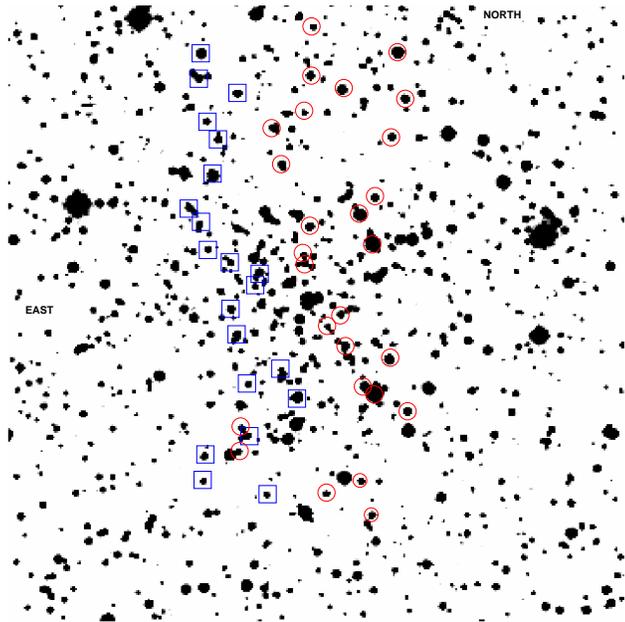}
  \caption{A 20 seconds $V$ exposure on Be~32 (9.4 $\times$ 9.4
  arcmin$^2$). The stars observed with MOS at DOLORES are
  indicated by red circles and blue boxes for Mask 1 and 2,
  respectively.}
\label{fig-map}
\end{figure}

\section{Observations and data reduction}

\begin{table}
 \centering
  \caption{Log of the photometric observations.}
  \begin{tabular}{cccrcc}
 \hline
Filter &  Date      &	  UT	 & Exp   & Airmass & Seeing \\
       & (dd mm yy) &		 & (sec) &	   & (\arcsec)\\
 \hline
\multicolumn{6}{c}{Berkeley 32: ~$\alpha_{2000}=06^h58^m07^s$,
$\delta_{2000}=+06^\circ 25'43''$} \\
 $B$ &   26 11 2000  &   03:39:38  &  600   & 1.081  &  2.34\\
 $B$ &   26 11 2000  &   03:36:59  &   40   & 1.082  &  2.20\\
 $B$ &   14 02 2004  &   20:41:59  &    5   & 1.210  &  1.19\\
 $V$ &   26 11 2000  &   02:47:20  &  480   & 1.118  &  1.79\\
 $V$ &   26 11 2000  &   02:43:45  &   20   & 1.122  &  1.65\\
 $V$ &   14 02 2004  &   20:37:08  &    2   & 1.223  &  1.38\\
 $I$ &   26 11 2000  &   03:03:45  &  480   & 1.100  &  1.98\\
 $I$ &   26 11 2000  &   03:01:26  &   20   & 1.102  &  2.06\\
 $I$ &   14 02 2004  &   20:39:33  &    1   & 1.216  &  0.96\\
\multicolumn{6}{c}{Control field:  ~$\alpha_{2000}=06^h57^m27^s$,
$\delta_{2000}=+06^\circ 08'26''$} \\
 $B$ &   26 11 2000  &   04:01:50  &  600   & 1.085  &  2.20\\
 $B$ &   26 11 2000  &   03:56:01  &  240   & 1.084  &  1.93\\
 $B$ &   26 11 2000  &   03:53:21  &   40   & 1.083  &  1.93\\
 $V$ &   26 11 2000  &   04:20:18  &  300   & 1.094  &  1.93\\
 $V$ &   26 11 2000  &   04:16:15  &  120   & 1.091  &  2.06\\
 $V$ &   26 11 2000  &   04:14:09  &   20   & 1.090  &  2.48\\
 $I$ &   26 11 2000  &   04:34:15  &  300   & 1.105  &  1.93\\
 $I$ &   26 11 2000  &   04:30:20  &  120   & 1.101  &  1.93\\
 $I$ &   26 11 2000  &   04:28:00  &   20   & 1.099  &  1.93\\
\hline
\end{tabular}
\label{tab-phot}
\end{table}

\subsection{Photometry}

Our data were acquired at the Italian Telescopio Nazionale Galileo, on the
Canary Islands, using  DOLORES (Device Optimized for the LOw RESolution), a
focal reducer capable of imaging and low resolution spectroscopy, on UT 2000
November 26, and UT 2004 February 14.  DOLORES mounts a 2k Loral thinned and
back-illuminated CCD, with scale of 0.275 arcsec pix$^{-1}$, and a field of
view of 9.4 $\times$ 9.4 arcmin$^2$.  We observed two fields, one centered on
the cluster (see Fig.~\ref{fig-map}) and a second one about 20 arcmin  away, to
determine the level of field star contamination.

For each field, Table \ref{tab-phot} lists the date of the observation together
with the filters used and other relevant information.  Seeing conditions were
better for the short exposures taken in 2004; however, since this is a rather
loose field, even modest seeing did not represent a problem for our photometry.
 We did not have any shutter timing problems, since they arise
only below 0.1 s with this instrument and even our shortest exposure time 
was 10 times longer.

The reduction procedure is the same adopted in the other papers of this series,
and details can be found e.g., in \cite{tosi04} or \cite{difabrizio05}. 
Standard IRAF
\footnote{IRAF is distributed by the National Optical Astronomical
Observatory, which are operated by the Association of Universities for
Research in Astronomy, under contract with the National Science
Foundation } routines were utilized for pre-reduction, and the IRAF version of
the DAOPHOT-{\sc ii} package (\citealt{stetson87}, \citealt{davis94}) was used
with a quadratically varying point spread function (PSF) to derive positions
and magnitudes for the stars. Output catalogues for each frame were aligned in
position and magnitude, and final (instrumental) magnitudes were  computed as
weighted averages of the individual values.

Aperture correction to bring the PSF magnitudes on the same scale of the
aperture magnitudes, i.e. the same of the standard stars, was computed for each
relevant frame.  The corrections range from about  $-0.15$ to $-0.35$ mag.

We observed the standard areas PG0231+051 and Rubin 149 \citep{landolt92} and
derived the calibration equations assuming the average extinction coefficients
for the site, as given in the site web pages (www.tng.iac.es:
$\kappa_B=0.25, \kappa_V=0.15, \kappa_I=0.07$): 

\[  B = b +0.0548 \times (b-v) + 1.464  ~~(rms=0.012) \]
\[  V = v -0.1448 \times (b-v) + 1.294  ~~(rms=0.027) \]
\[  V = v -0.0933 \times (v-i) + 1.270  ~~(rms=0.021) \]
\[  I = i +0.0228 \times (v-i) + 0.846  ~~(rms=0.021) \]

where $b$, $v$, $i$, are the aperture-corrected instrumental magnitudes, after
further correction for extinction and for exposure time, and $B$, $V$, $I$ are
the output magnitudes, calibrated to the Johnson-Cousins standard system.

Finally, we determined our completeness level in each band using extensive
artificial stars experiments (see \citealt{tosi04}), whose results are given
in Table 2. %\ref{tab-compl}

We also checked out photometry against  the studies of \cite{km91} and
\cite{rs01}, finding a very good agreement.   The stars in common, 
about 800 with \cite{km91} and 1700 with \cite{rs01}, respectively, cover
the whole interval in magnitude and colours. They have a mean
difference (our photometry minus literature) of $\Delta B= -0.003$
($\sigma$=0.070), $\Delta V= -0.015$ ($\sigma$=0.061), for \cite{km91} and
$\Delta I= -0.011$ ($\sigma$=0.101), $\Delta V= +0.015$ ($\sigma$=0.072) for
\cite{rs01}  (i.e., we are intermediate between the two photometries in
$V$). No trends with magnitude or colour are present.

\begin{table}
\begin{center}
\caption{
Completeness level for the central and external fields; mag is
the calibrated magnitude ($B, V$ or $I$, calibrated with the equations given
in the text, and assuming $b-v$ and $v-i$ = 1).}
\begin{tabular}{ccccccc}
\hline
mag & $c_B$ & $c_V$ & $c_I$ & $c_B$ & $c_V$ & $c_I$ \\
 & \multicolumn{3}{c}{(central field)}
 & \multicolumn{3}{c}{(external field)} \\
\hline
  16.00 & 1.00 & 1.00 & 1.00  & 1.00 & 1.00 & 1.00 \\
  16.50 & 0.97 & 0.95 & 0.92  & 1.00 & 0.99 & 0.95 \\
  17.00 & 0.89 & 0.87 & 0.88  & 0.99 & 0.98 & 0.94 \\
  17.50 & 0.87 & 0.85 & 0.85  & 0.97 & 0.97 & 0.92 \\
  18.00 & 0.80 & 0.81 & 0.78  & 0.97 & 0.94 & 0.87 \\
  18.50 & 0.74 & 0.69 & 0.68  & 0.96 & 0.93 & 0.84 \\
  19.00 & 0.67 & 0.59 & 0.54  & 0.93 & 0.93 & 0.73 \\
  19.50 & 0.52 & 0.45 & 0.37  & 0.91 & 0.90 & 0.52 \\
  20.00 & 0.33 & 0.28 & 0.21  & 0.89 & 0.86 & 0.29 \\
  20.50 & 0.16 & 0.10 & 0.09  & 0.85 & 0.78 & 0.11 \\
  21.00 & 0.07 & 0.05 & 0.03  & 0.69 & 0.58 & 0.04 \\
  21.50 & 0.02 & 0.01 & 0.01  & 0.42 & 0.32 & 0.01 \\
  22.00 & 0.00 & 0.00 & 0.00  & 0.22 & 0.15 & 0.00 \\
  22.50 & 0.00 & 0.00 & 0.00  & 0.07 & 0.05 & 0.00 \\ 
\hline
\end{tabular}
\end{center}
\label{tab-compl}
\end{table}

\subsection{Spectroscopy}

\begin{figure}
 \includegraphics[width=84mm, bb=70 160 570 700 , clip]{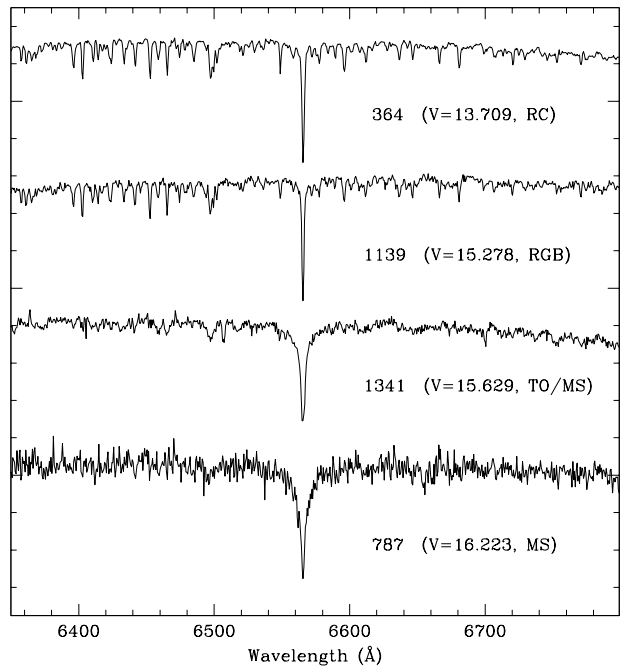}  
  \caption{A few representative spectra, shifted for clarity; from top to
  bottom we show a RC, a RGB and two TO/MS stars}
\label{fig-spe}
\end{figure}

The targets for the spectroscopic observations were selected from the
preliminary photometry of images taken during our first run, trying to sample
the relevant evolutionary phases: red giant branch (RGB), red clump (RC), upper
main sequence (MS) and main-sequence  Turn-Off (TO) point. We also added two
stars near the RGB tip, taken from published catalogues (stars 2 and 4 in
\citealt{km91}). These stars are too bright for the exposure times of our
first  photometric images and only one of them was later completely recovered
in our second run. A few  probable field stars, based on the preliminary
colour-magnitude diagram, were also included as a check.
The spectra were acquired using the  Multi Object Spectroscopy (MOS) facility
of DOLORES on UT February 14, 16, 17   2004; 4 exposures were take with the
first mask,  and 3 with the second one (see Table 3), using the VPH grism
centered on H$\alpha$ and a  1\farcs1 slit   (resolution 1.36 \AA, or 
\rr~$\simeq 4800$). 
 The full wavelength range of our spectra is 680 \AA, but in
Fig. \ref{fig-spe} only a region centred on H$\alpha$ is shown. 
The configuration is the same already used for Berkeley 29
\citep{bht05}, as is the reduction procedure. 

Given the much better S/N of most
of  the present spectra  (see Fig. \ref{fig-spe} for some representative cases
with S/N about 25, 75, 125, 150 from bottom to top), a different strategy was
instead adopted to derive the radial velocity (RV). To this end we used the
task {\em rvidlines} in IRAF with a list of about 35 lines, mostly of iron, and
comprising H$\alpha$, the only line always visible in all spectra, even at very
low S/N and/or high temperature.  The formal error of the velocities measured by
{\em rvidlines} is of about 5 km s$^{-1}$. The observed velocities were
corrected to heliocentric, and averaged (see Section~4).

\begin{table}
\begin{center}
\caption{Log of the spectroscopic observations}
\begin{tabular}{|ccccc|}
\hline
 MASK    &   Date    &  UT	   &  Exp  & Airmass	\\
         &(dd mm yy) & (beginning) & (sec) &		 \\
\hline
mask1\_a &  14 02 2004 &  21:12:01 & 1200  &   1.146 \\
mask1\_b &  14 02 2004 &  21:40:38 & 1200  &   1.106 \\
mask1\_c &  14 02 2004 &  22:06:15 & 1800  &   1.086 \\
mask1\_d &  14 02 2004 &  22:42:54 & 1800  &   1.081 \\
mask2\_a &  16 02 2004 &  02:11:49 & 1200  &   1.833 \\
mask2\_b &  16 02 2004 &  02:37:07 & 1200  &   2.159 \\
mask2\_c &  17 02 2004 &  02:11:09 & 1200  &   1.868 \\
\hline
\end{tabular}
\end{center}
\label{tab-spec}
\end{table}

\section{The Colour-Magnitude Diagrams}

\begin{figure*}
 \includegraphics[bb=40 150 680 680 , clip]{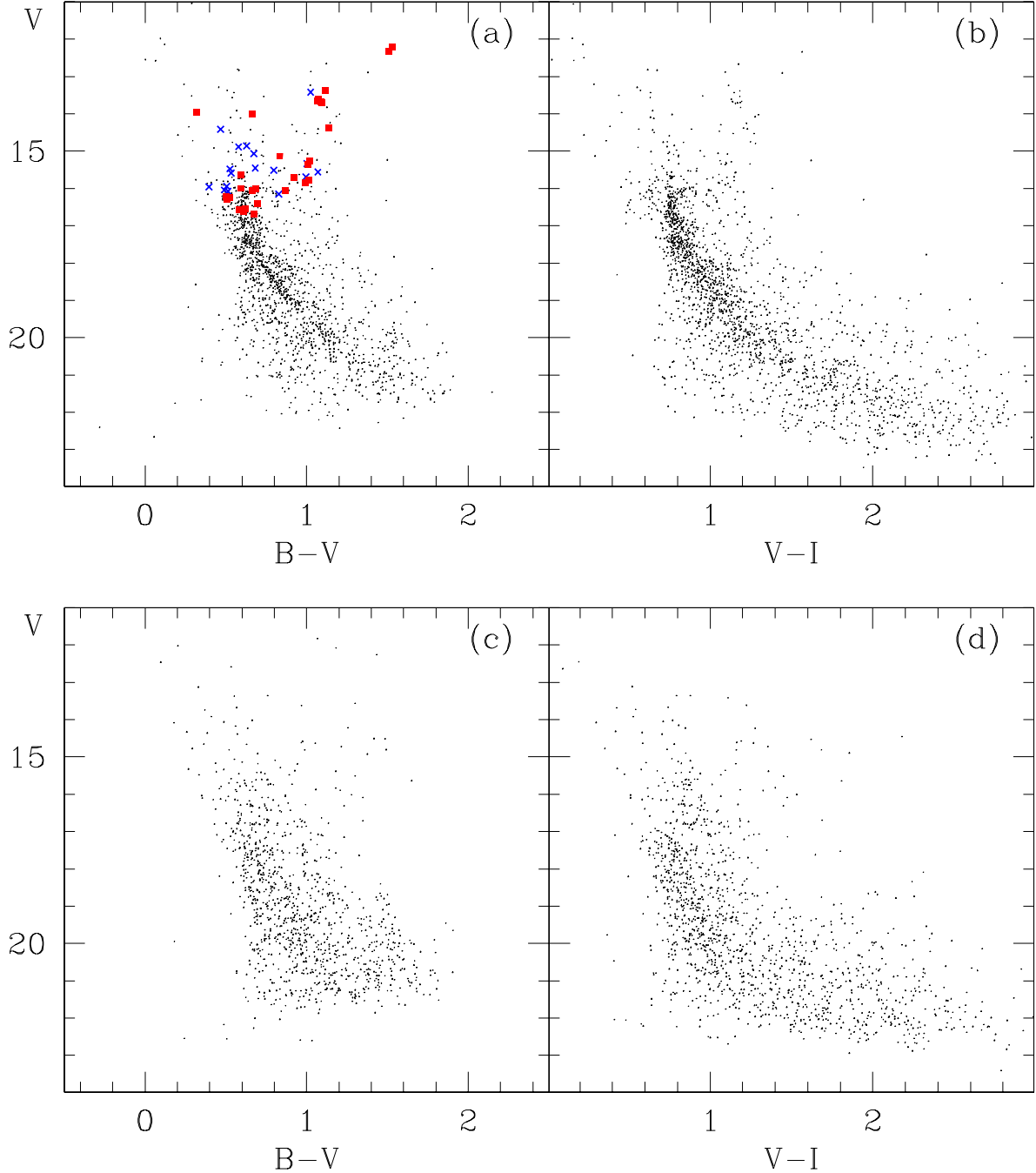}  
  \caption{(a) and (b) CMD for Be~32 based on our central field. 
   The main sequence and red
  clump of the cluster are easily discernible.
  We indicate, only in (a) for clarity, the
  position in the CMD of the stars observed with DOLORES; (red) filled
  squares and (blue) crosses indicate members and non members,
  respectively, on the basis of their RVs (see Sect. 4).
  (c) and (d) The CMD for the control field. 
    }
\label{fig-cmd}
\end{figure*}

The final photometric cluster catalogue\footnote{
The catalog 
can be obtained in electronic form  through the BDA \citep{mermio95} at
http://www.univie.ac.at/webda/new.html} 
includes 2031 objects;  1541 of them are identified in  all 3 filters, 1576 at
least in  $B$ and $V$ and 1996 at least in   $V$ and $I$.  Similarly, the
external field catalogue has 1398 objects (1031 of them with magnitudes in all
3 bands, 1069 at least in $B$ and $V$, and 1348 in $V$ and $I$). The
corresponding CMDs are shown in Fig. \ref{fig-cmd}, where we also anticipate
results on membership, described in Sect. 4. Pixel coordinates for all Be~32
stars were transformed to equatorial coordinates using  software written by P.
Montegriffo; we used a linear relation, and residuals have a rms of about 0.15
arcsec in both coordinates. 

\begin{figure*}
 \includegraphics[bb=40 410 680 680, clip]{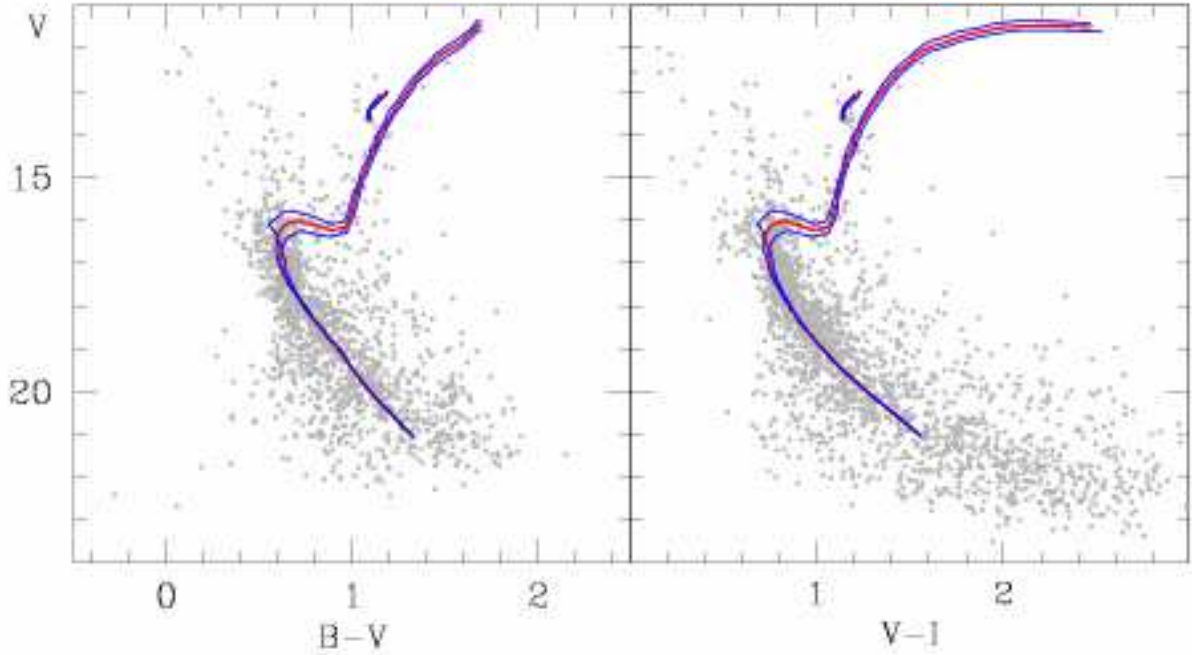}  
  \caption{Best fit isochrones for Be~32, obtained with Z = 0.008, \ebv = 0.10,
   \mmm = 12.48, age = 6.3 Gyr. Also shown are the two bracketing ages,
   5.0 and 7.9 Gyr, which produce a worse fit.}
\label{fig-iso}
\end{figure*}

\begin{figure*}
 \begin{center}
 \includegraphics [bb=20 460 680 680, clip, scale=0.95]{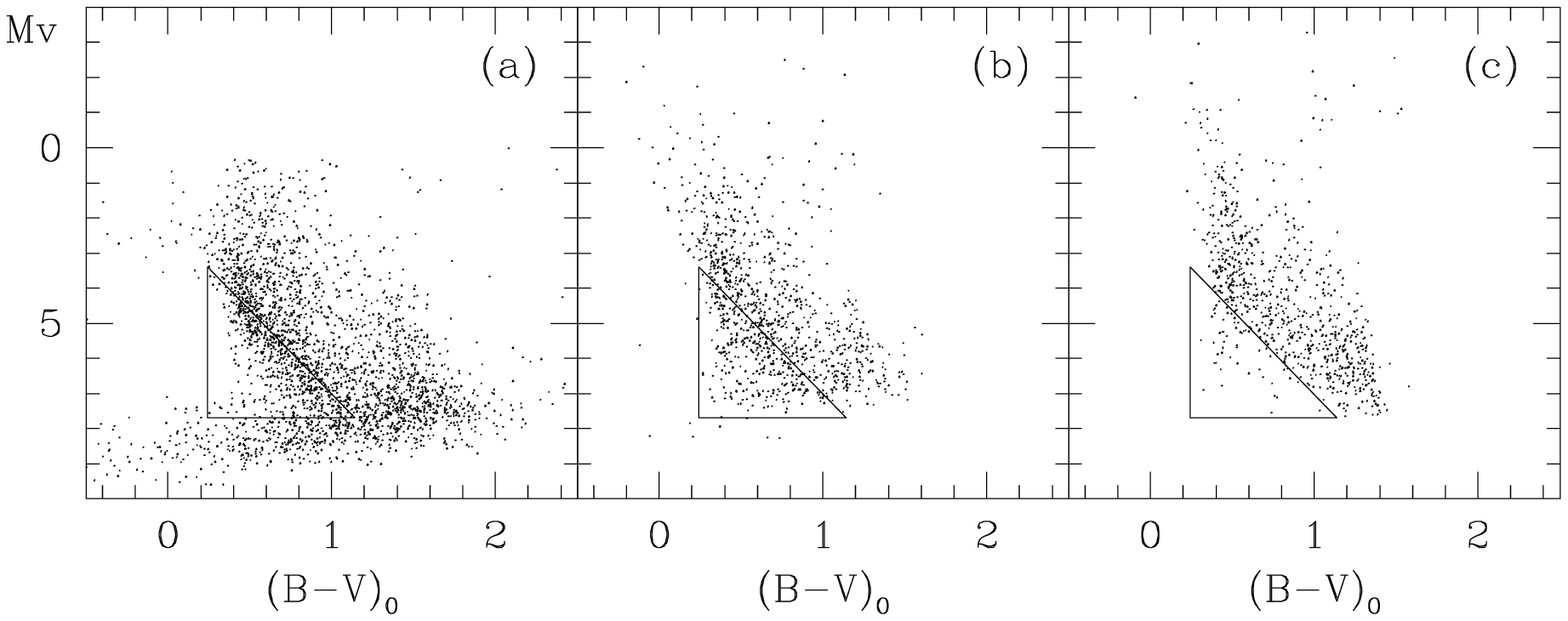}
  \caption{ The absolute magnitude versus dereddened colour
  diagrams for: 
  (a) CMa [using \mmm=14.5, \ebv=0.10 from \citep{bellazzini04}], plotting only
  the fraction of stars implied by the different areas; 
  (b) our external field [using \mmm=13.4, \ebv =0.3: this reddening value
  compares very well with the one derived in the \citealt{sfd98} maps, which is
  0.295 mag in this direction];
  (c) the Besan\c con model, computed on an equal-area field, shifted as our
  field and adopting
  the completeness of our data. The lines on the blue of the MS indicate the
  regions of the CMDs we are comparing.}
\label{fig-ring}
\end{center}								     
\end{figure*}

  Some disc star contamination is noticed in the cluster CMD,
  yet we are able to locate the TO and RC at								     
$V \simeq 16.0$, $B-V \simeq 0.6$, $V-I \simeq 0.7$,  and
$V \simeq 13.7$, $B-V \simeq 1.1$, $V-I \simeq 1.2$, 
respectively.
Most stars falling out of the main features are contaminating field
objects, but there are a few ones, brighter than the TO and bluer than the RGB,
that may  instead be blue stragglers; they are worth of further
examination.   Furthermore, an accurate measurement of the RC mean magnitude
is  obtained using  the 4 RC stars with spectroscopically confirmed membership
(see next Section). We derive  $\langle V \rangle = 13.66$ (r.m.s.~$= 0.05$)
and $\langle I \rangle = 12.53$ (r.m.s.~$= 0.04$), where the errors are the
standard deviations of the measurements.

Be~32 appears to be old; a first measure of its age can be obtained through
morphological age indicators, like the Morphological Age Index 
\citep[MAI:][]{jp94}, based on the difference in magnitude between the TO and
the RC  ($\delta V$), or the Morphological Age Ratio  \citep[MAR:][]{att85}, 
which also takes into account the difference in colour between the two points.
From our data we obtain $\delta V \simeq$ 2.3 and $\delta (B-V) \simeq$ 0.65;
these figures have an attached uncertainty of about 0.1 mag, since the actual
TO and RC points are difficult to locate in a diagram contaminated by field
stars (even if we were helped by the membership information of selected stars).
They translate into ages of about 6 Gyr using the MAI, and between 5 and 7 Gyr
using the original \citep{att85} and the revised \citep{att89} calibrations of
the  MAR, respectively.  Our measure is in good agreement with the
age derived both from morphological age parameters and
isochrone fitting by \cite{rs01}. 
This has to be taken  as a first indication of the
cluster age, but is substantiated also by the results by \cite{salaris04}. They
calibrated a similar method, that makes also use of information on the
metallicity; using the literature values of $\delta V$ = 2.4 and
[Fe/H]=$-$0.50, they derived a very similar age of 5.91 Gyr.

We have fitted our data with isochrones, using the models of \cite{bertelli94},
since they come from the same set of evolutionary tracks that we have already
successfully used  in our works (\citealt{bt06} and references therein).  We
tried three different values of metallicity, Z = 0.02 (i.e., solar abundance), 
Z = 0.008 (corresponding to [Fe/H] $\simeq -0.4$) and Z = 0.004 (corresponding
to [Fe/H] $\simeq -0.7$).  The first and last sets of isochrones did not fit
well our data, while isochrones with Z = 0.008 provided a good fit. This is in
agreement with the most recent determination of the cluster metallicity [Fe/H]
= $-$0.50 \citep{friel02}.\footnote{
Preliminary analysis of the FLAMES-UVES spectra (see next Section) also
indicate a metallicity similar to Z = 0.008}
The results of the best fit, obtained with \ebv = 0.10, \mmm =
12.48 and age = 6.3 Gyr (and Z=0.008), are shown in Fig.~\ref{fig-iso}.

An independent estimate of the distance has been derived by an empirical
comparison of the RC in Be~32 with the RC level in open clusters
with similar properties and well established distance
\citep[see][ for details]{bht05}.
We chose as template cluster  Mel\,66, an open cluster for which
\citet{saraj99} reports  [Fe/H]~$= -0.35$ and an age 4.5 Gyr, hence quite
similar to Be~32.
Since the metallicities of Be~32 and Mel\,66 are indistinguishable within the
uncertainties, there is no need to correct for the dependence of the RC
luminosity on metallicity. Instead, we applied  a small correction for
population effects to the $V$ and $I$ luminosity of the RC assuming that Be~32
is 2 Gyr older than Mel\,66, and using the calculations  of
\citet{girasala01}.  The correction is of the order 0.07 mag (for $Z=0.008$)
and implies luminosities $M_V = 0.81$ and $M_I = -0.27$ for the red clump of
Be~32.

With this population correction, and assuming a reddening  \ebv~$ = 0.10 \pm
0.05$ from our isochrone fit,  we obtained an extinction-corrected distance
modulus
\mmm$= 12.54 \pm 0.16$ using $V_{\rm RC}$ and 
\mmm$= 12.62 \pm 0.09$ from $I_{\rm RC}$, 
where the errors mainly reflect the uncertainty on the
reddening. 

These results do not differ very much from the literature ones (see
Introduction). We defer any further comparison until  a refinement of our results
is obtained using a synthetic CMD technique.

The CMD of the external field appears to be composed by (at least) two
populations: a broad young main sequence of field stars,  crossing diagonally
the diagram, and a second more concentrated sequence  fainter and bluer than
the Be~32  main sequence.  The latter does not seem, at first sight,
completely ascribable to the "normal" disc population, as described, e.g., by
the Besan\c con Galactic model by \cite{robin03}.
A thorough discussion is beyond the purposes of the present paper (and the
possibilities of our data), but we consider  it useful to further check whether
we could be seeing some extra-disc component. Given the position of our field
($l=208^\circ$, $b=+4.1^\circ$) and the distances involved we cannot have
intercepted directly  the Monoceros Ring (e.g., \citealt{newberg02},
\citealt{ibata03}). This feature of the Galactic disc has also been
suggested (\citealt{martin04}) to be
associated to an over density in star counts that could  represent the remnant
of a dwarf galaxy (Canis Major, CMa). While the existence of the Ring itself
is not disputed, the reality of CMa has however been challenged,  for instance, by
\cite{momany04} who attribute  the over density to the Galactic Warp.  
We present in Fig. ~\ref{fig-ring} a comparison
of our data [Fig. ~\ref{fig-ring}(b)] to what is thought to be the CMD of
CMa  [\citep{bellazzini04}, Fig. ~\ref{fig-ring}(a)], and to the modeled
disc population [Fig. ~\ref{fig-ring}(c)].
All CMDs  have been scaled to absolute magnitudes  and
dereddened colours by correcting for the values given in the caption. 
The apparently better agreement between the CMDs in panels (a) and (b) 
suggests that  we might be seeing a portion of the
disrupting CMa \citep[see also ][]{bellazzini06}.

To say something more definitive, further investigations would be required,
involving e.g. precise radial velocities of lower MS stars, to be
compared to the ones that are being derived in the direction of CMa and of
several Ring positions.  As indicated in the
next section, our field stars have an average RV of 
about 43 km~s$^{-1}$ (with an rms of 28 km~s$^{-1}$), similar to that
of the disc stars in the same direction, as deduced from the Besan\c con model
(33 km~s$^{-1}$, with an rms of 28). However, our velocities refer  
to stars much brighter than those of the faint blue CMD feature which might
correspond to extra-disc components. The bright stars of our sample are most
likely field objects absolutely compatible with the model in Fig. 
~\ref{fig-ring}(c).

\begin{figure}
 \includegraphics[width=84mm, bb=40 150 330 680, clip]{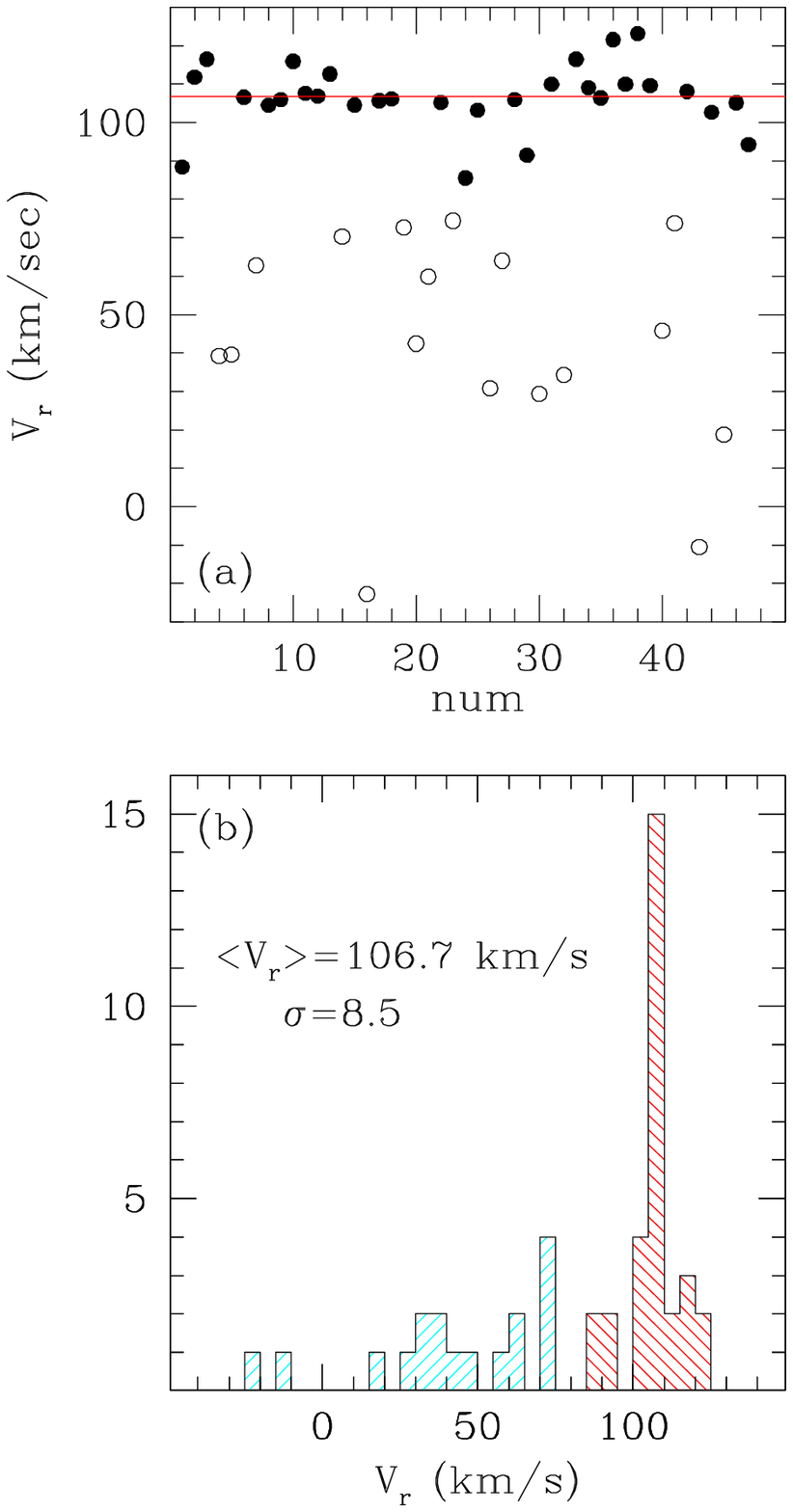} 
  \caption{(a) Heliocentric radial velocities for all spectra: filled circles
  indicate cluster members (RV $>$ 85 km s$^{-1}$, and open circles 
   are field stars.
  (b) Histogram of the RVs; red and light blue shadings of different slanting
  indicate the cluster and field stars distributions, respectively.}
\label{fig-vr}
\end{figure}

\begin{table*}
\begin{minipage}{126mm}
\caption{Radial velocities of stars in Berkeley~32. 
The first column shows the ID number in our photometry, KM
and RS indicate the corresponding identifier in Kaluzny \& Mazur (1991) and 
Richtler \& Sagar (2001)
respectively; RA and Dec are the equatorial coordinates at J2000; RV is the
heliocentric radial velocity and $\sigma$ is the standard deviation from the
mean RV; RV$_{\rm SFJ}$ is the radial velocity in Scott et al. (1995).
The magnitudes come from our catalogue (except
for star 2689 - which we measured only in $B$ and $I$ - where  $B, V$ come
from  \citealt{km91}). The last
column indicates the membership  status (C for cluster member, F for field
stars). The ordering reflects the position of the apertures in each of the two
masks. Star 1502 was observed in both masks, and the RV given here is the
average value.}
\begin{tabular}{rrrccrrrcccc}
\hline
 Star &  KM & RS  & RA(2000) &Dec(2000) & RV  &$\sigma$ &RV$_{\rm SFJ}$                & $B$ & $V$  & $I$ & M\\
      &     &     & (h:m:s)  & (d:m:s)  & \hspace{10pt}(km s$^{-1}$) & (km s$^{-1}$) & (km s$^{-1}$)&   &    &   &  \\
\hline
\multicolumn{12}{c}{ Mask 1 } \\
  345 &   \nodata &  563 & 6:58:06.351 &  6:22:45.72&  88.44  & 9.64 &     & 16.804 & 16.298&  15.662 & C \\
   18 & 351 & 1720 & 6:58:04.037 &  6:22:58.04& 111.84  & 6.00 &     & 16.728 & 16.066&  15.243 & C \\
  408 & 161 &  525 & 6:58:12.034 &  6:23:26.65& 116.47  & 2.86 &     & 17.166 & 16.581&  15.811 & C \\
  465 & 116 &  488 & 6:58:11.861 &  6:23:49.03&  39.29  & 5.43 &     & 16.975 & 16.148&  15.241 & F \\
  500 &  49 & 1243 & 6:58:01.030 &  6:24:06.63&  39.54  & 1.29 &     & 16.337 & 15.334&  14.275 & F \\
 2689 &	  2 &	 \nodata & 6:58:03.398 &  6:26:49.89& 106.57  & 4.46 & 102 & 13.733 & 12.201&	 \nodata  & C \\  
  550 &  26 &  446 & 6:58:03.840 &  6:24:30.16&  62.76  & 4.77 &     & 14.879 & 14.412&  13.849 & F \\
  605 &  27 &  416 & 6:58:02.269 &  6:24:56.76& 104.48  & 1.75 & 109 & 15.511 & 14.372&  13.177 & C \\
 1341 &  57 & 1179 & 6:58:05.104 &  6:25:10.52& 105.95  & 7.00 &     & 16.224 & 15.629&  14.888 & C \\
 1523 & 236 &	 \nodata & 6:58:06.244 &  6:25:28.58& 115.92  & 4.50 &     & 17.359 & 16.682&  15.928 & C \\
  698 &  75 &  357 & 6:58:05.417 &  6:25:40.24& 107.61  & 5.18 &     & 16.840 & 15.846&  14.798 & C \\
  737 &  16 & 1685 & 6:58:06.945 &  6:25:56.45& 106.91  & 3.07 &  93 & 14.672 & 13.597&  12.478 & C \\
  787 &  19 &  318 & 6:58:03.095 &  6:26:16.08& 112.61  & 3.39 &  87 & 14.800 & 13.709&  12.564 & C \\
% 820 & 102 &  303 & 6:58:07.792 &  6:26:29.61&    0    &      & & 16.625 & 16.062&  15.429 &  \\
  110 &  82 &  292 & 6:58:07.836 &  6:26:37.71&  70.30  & 4.60 &     & 16.359 & 15.962&  15.509 & F \\
 1556 &	  4 &	 \nodata & 6:58:03.182 &  6:24:22.32& 104.55  & 4.04 &  98 & 13.830 & 12.321&  10.773 & C \\
  895 &  39 & 1650 & 6:58:07.459 &  6:27:05.82& -22.78  & 7.70 &     & 15.739 & 15.065&  14.326 & F \\
 1393 &  12 & 1948 & 6:58:04.223 &  6:27:17.15& 105.69  & 4.87 &     & 14.492 & 13.375&  12.203 & C \\
 1535 &  62 & 1642 & 6:58:03.247 &  6:27:34.16& 106.09  & 3.77 &     & 16.620 & 15.699&  14.452 & C \\
 1502 &  58 &  963 & 6:58:09.303 &  6:28:05.80&  72.65  & 5.72 &     & 16.120 & 15.588&  14.841 & F \\
  169 &  51 & 1629 & 6:58:02.148 &  6:28:33.11&  42.46  & 5.82 &     & 16.002 & 15.476&  14.750 & F \\
  173 &  61 & 1626 & 6:58:09.894 &  6:28:41.84&  59.85  & 2.33 &     & 16.690 & 15.693&  14.594 & F \\
 1079 & 167 &  911 & 6:58:07.857 &  6:28:56.51& 105.19  & 2.58 &     & 17.164 & 16.544&  15.766 & C \\
 1101 &  35 & 1616 & 6:58:01.273 &  6:29:10.52&  74.43  & 3.88 &     & 15.485 & 14.855&  14.068 & F \\
  186 &  22 & 2220 & 6:58:05.383 &  6:29:19.28&  85.57  &13.81 &     & 14.290 & 13.972&  13.506 & C \\
 1139 &  44 & 1866 & 6:58:07.474 &  6:29:32.61& 103.20  & 2.22 &     & 16.299 & 15.278&  14.153 & C \\
 1183 &   \nodata & 1851 & 6:58:01.771 &  6:29:55.30&  30.77  & 4.47 &     & 14.450 & 13.425&  12.282 & F \\
\multicolumn{12}{c}{ Mask 2 } \\ 			   
  343 &   \nodata  & 565 & 6:58:10.213 &  6:22:44.68&  64.03  & 4.62 &     & 16.585 & 16.071&  15.424 & F \\
  364 &   \nodata  &2269 & 6:58:14.334 &  6:22:57.83& 105.93  & 5.04 &     & 16.726 & 16.223&  15.573 & C \\
 1574 & 178  &	 \nodata & 6:58:14.227 &  6:23:22.25&  91.45  &12.17 &     & 17.104 & 16.410&  15.605 & C \\
  445 &  53  &1274 & 6:58:11.509 &  6:23:41.66&  29.41  & 2.15 &     & 16.313 & 15.515&  14.654 & F \\
  533 &  17  & 456 & 6:58:08.244 &  6:24:19.55& 109.97  & 1.18 & 115 & 14.754 & 13.667&  12.540 & C \\
   57 &  81  & 442 & 6:58:11.428 &  6:24:32.32&  34.31  & 0.63 &     & 16.449 & 15.945&  15.314 & F \\
  577 & 172  &1715 & 6:58:09.296 &  6:24:46.16& 116.44  & 8.24 &     & 17.245 & 16.633&  15.936 & C \\
  666 &  71  &1702 & 6:58:12.186 &  6:25:20.83& 109.10  & 1.73 &     & 16.794 & 15.783&  14.718 & C \\
  710 &  41  &1691 & 6:58:12.586 &  6:25:44.94& 106.39  & 3.43 &     & 15.968 & 15.135&  14.171 & C \\
   97 & 129  &1680 & 6:58:11.008 &  6:26:06.94& 121.55  & 5.11 &     & 16.778 & 16.254&  15.650 & C \\
  799 &  50  &2239 & 6:58:10.734 &  6:26:21.16& 109.96  & 2.55 &     & 16.354 & 15.348&  14.263 & C \\
  822 &  91  & 302 & 6:58:12.609 &  6:26:30.27& 123.18  &10.31 &     & 16.605 & 16.010&  15.276 & C \\
  113 &  97  &1661 & 6:58:14.080 &  6:26:43.09& 109.65  & 4.59 &     & 16.930 & 16.060&  15.056 & C \\
  898 &  99  &1027 & 6:58:14.622 &  6:27:08.11&  45.79  & 9.66 &     & 16.533 & 16.041&  15.412 & F \\
 1394 &  59  & 238 & 6:58:15.315 &  6:27:23.25&  73.75  & 1.25 &     & 16.633 & 15.564&  14.415 & F \\
  997 &  18  & 974 & 6:58:13.758 &  6:27:54.78& 108.07  & 4.12 & 105 & 14.730 & 13.661&  12.534 & C \\
  166 &  52  &1895 & 6:58:13.477 &  6:28:29.70& -10.49  & 3.42 &     & 16.138 & 15.456&  14.545 & F \\
 1070 &  96  & 157 & 6:58:14.201 &  6:28:47.59& 102.63  & 3.22 &     & 16.697 & 16.012&  15.185 & C \\
  183 &  36  &2221 & 6:58:12.193 &  6:29:14.95&  18.79  & 5.15 &     & 15.463 & 14.884&  14.142 & F \\
 1135 &  64  & 115 & 6:58:14.727 &  6:29:29.77& 105.12  & 3.27 &     & 16.621 & 15.701&  14.689 & C \\
  199 &   \nodata  &1605 & 6:58:14.580 &  6:29:53.84&  94.30  & 4.85 &     & 14.683 & 14.018&  13.252 & C \\
\hline							   
\end{tabular}
%\end{center}
\end{minipage}
\label{tab-vr}
\end{table*}

\section{Radial velocities: membership}
\label{sec:memberstars}

We obtained RVs for 48 of the 49 objects observed; one star near the TO had
very low S/N spectra and we did not determine its RV. We averaged the 4 or 3
values for the stars in Mask1 and Mask2, respectively, after applying the
heliocentric corrections. These average values are given in Table \ref{tab-vr},
together  with their r.m.s. The RV for star 1502 was determined independently
for both masks (heliocentric RV= 66.92 and 78.37 km s$^{-1}$ for Mask1 and
Mask2, respectively), and we give  here the average of the two values.   Given
the precision expected on the basis of the resolving power, the internal 
errors of the RV measures, the standard deviations of the means, and this last
result, we may safely attach an average error of $\pm$ 5  km s$^{-1}$ to our
velocities.

Results of our RV study  are summarized in  Fig. \ref{fig-vr}. Panel (a) 
shows all the measured RVs, and the derived mean cluster
velocity, after elimination of all probable field objects (the limit at 
 RV = 85
km s$^{-1}$ is equivalent to a 2.5$\sigma$  clipping). Panel (b) is a histogram
indicating the mean RV and the tail of field stars.  The average cluster
velocity derived from our data is  $\langle$RV$\rangle$ = 106.7 km s$^{-1}$ (rms
8.5  km s$^{-1}$, 31 stars).

We have six stars in common with \cite{scott95}, who derived RVs with a
precision of about 10  km s$^{-1}$. The RVs are in good  agreement (see Table
\ref{tab-vr}), with an average difference (MOS minus \citealt{scott95}) of
about 6 km s$^{-1}$, that becomes about 3 km s$^{-1}$ excluding the most
discrepant star.

We have used the membership information based on our RVs to select RGB and RC
stars to be observed with FLAMES-UVES (R $\simeq$ 40000) at the ESO Very Large
Telescope in January 2005. The preliminary UVES measurements of the RVs of six
stars in common are  in very good agreement with the present values (P.
Sestito, private communication): $\Delta$RV(MOS-UVES)  = 2.53 ($\sigma$ = 4.92)
km s$^{-1}$, with the difference being dominated by 1 star (without it, we have
$\Delta$RV = 0.80, $\sigma$ = 2.78 km s$^{-1}$).

\section{Summary}

We have presented a photometric and spectroscopic study of stars in the 
old open cluster Berkeley~32. The main results are the following:

\begin{enumerate}
\item Analysis of the cluster CMDs confirms that this is an old,
nearby, rather metal-poor cluster (age $\simeq$ 6 Gyr, \mmm = 12.48, \ebv =
0.10, Z = 0.008). 

\item An analysis of the mean luminosity of RC stars 
provides an independent confirmation of this distance to Be~32, 
yielding \mmm~$\simeq 12.6 \pm 0.1$.

\item The CMD of the comparison field  looks similar to the one
of CMa, i.e. of the proposed originator of the Anticentre Ring.
 
\item Membership of 48 stars in the cluster directions has been determined on
the basis of their RV. 

\item The cluster average RV has been measured:  106.7 km s$^{-1}$, rms
of 8.5 km s$^{-1}$, based on 31 member stars.

\end{enumerate}

Further work on this interesting cluster is foreseen, starting from the
present results: for instance, the information on 
membership will be useful when analyzing Be~32 with the synthetic CMD 
technique and the $B,V$ photometry in deriving atmospheric parameters
of stars observed with high resolution spectroscopy.

%------------------------------------------------------------------------
\section*{Acknowledgments}
We thank M. Bellazzini and S. Zaggia for valuable discussions on CMa  and P. Montegriffo for
his very useful software. We gratefully acknowledge the use of the BDA by J.-C.
Mermilliod. This project has been partially supported by the Italian MIUR,
under PRIN 2003029437.

\bsp

\label{lastpage}

\end{document}